\documentclass[aps,prl,twocolumn,superscriptaddress,showpacs]{revtex4}

\usepackage{graphicx}
\usepackage{tabularx}
\begin{document}

\title{Large Splitting of the Cyclotron-Resonance Line in AlGaN/GaN Heterostructures }

\author{S. Syed}
\affiliation{Department of Applied Physics and Applied
Mathematics, Columbia University, New York, New York 10027}
\author{M.J. Manfra}
\affiliation{Bell Laboratories, Lucent Technologies, Murray Hill,
NJ 07974}
\author{Y.J. Wang}
\affiliation{National High Magnetic Field Laboratory, Florida
State University, Tallahassee, FL 32306}
\author{H.L. Stormer}\affiliation{Department of Applied Physics and Applied
Mathematics, Columbia University, New York, New York
10027}\affiliation{Bell Laboratories, Lucent Technologies, Murray
Hill, NJ 07974}\affiliation{Department of Physics, Columbia
University, New York, New York 10027}
\author{R.J. Molnar}
\affiliation{MIT Lincoln Laboratory, Lexington, MA 02420-0122}

\date{\today}

\begin{abstract}
Cyclotron-resonance (CR) measurements on two-dimensional (2D)
electrons in AlGaN/GaN heterojunctions reveal large splittings (up
to 2 meV) of the CR line for all investigated densities, $n_{2D}$,
from $1$ to $4\times10^{12}$cm$^{-2}$ over wide ranges of magnetic
field. The features resemble a level anti-crossing and imply a
strong interaction with an unknown excitation of the solid. The
critical energy of the splitting varies from 5 to 12 meV and as
$\sqrt{n_{2D}}$ . The phenomenon resembles data from AlGaAs/GaAs
whose origin remains unresolved. It highlights a lack of basic
understanding of a very elementary resonance in solids.
\end{abstract}
 \maketitle

Ever since its first demonstration in 1953 \cite{Kittel53},
cyclotron resonance (CR) has become the most widely used technique
to determine effective masses of carriers in semiconductors and
their heterojunctions. In the presence of an external magnetic
field, $B$, electrons of charge $e$ are set into cyclotron motion
whose frequency is given by $\omega_c=eB/m^*c$. An RF or far
infra-red source is swept which generates a single absorption line
at $\omega_c$, from which the effective mass $m^*$ is deduced. CR
has also been instrumental in detecting other excitations of the
solid via their interaction with CR. They often take on the shape
of a level anti-crossing from which the energy of the new
excitation and its coupling to CR can be inferred. These are
well-established, textbook phenomena, which make CR one of the
best-understood tools in solid state research \cite{McCombe1994}.
Yet, in our experiments on the comparatively new AlGaN/GaN
heterostructures we observe huge splittings in the CR line,
indicating a strong interaction with an excitation whose origin
remains unknown. Our results provide a fundamental challenge to
our apparently thorough understanding of CR in solids.

For AlGaN/GaN heterostructures there exist only few CR data
\cite{Wang96,Knap96,Knap97,ZFLi02}. Our study of CR in
two-dimensional electron systems (2DESs) in a sequence of
high-quality AlGaN/GaN structures with a wide range of carrier
densities ($1$ - $4\times10^{12}$cm$^{-2}$) reveals splittings in
the CR line reminiscent of level anti-crossings, reaching $20\%$
of the resonance energy. While their origin is not established,
they resemble splittings seen previously in the CR of AlGaAs/GaAs
and Si systems. Our data do not support a universal
\textit{ad-hoc} model put forward in the AlGaAs/GaAs work and
points to the lack of a theoretical understanding of this
phenomenon, now seen in three different 2D systems.

Our AlGaN/GaN samples are grown by plasma-assisted MBE on thick
GaN ($\sim$15$\mu$m) templates prepared by hydride vapor phase
epitaxy (HVPE) on the $[0001]$ face of sapphire. The use of thick
HVPE grown templates is essential for achieving low threading
dislocation densities in the GaN buffer region.  The reduction of
the threading dislocation density has been shown to improve low
temperature mobility at low electron densities.  The typical MBE
layer sequence consists of $400$nm of GaN, followed by a layer of
$25$ to $50$nm Al$_x$Ga$_{1-x}$N, which is capped by $3$nm of GaN.
In contrast to the AlGaAs/GaAs system, nitride heterostructures
are not modulation doped.   The creation of a 2DES rather
originates from strong spontaneous and piezoelectric fields
arising at the heterointerface \cite{Ambacher99,Smorchkova99}. The
Al mole fraction and the thickness of the barrier layer control
the $2$D electron density. In our samples the Al content varies
from $3$ to $10\%$, yielding the parameters listed in Table I.
Samples with $n_{2D}\sim 1 - 3\times10^{12}$cm$^{-2}$ show the
integer \cite{Manfra00} and the fractional quantum Hall effects
\cite{Manfra02}, further attesting to the high quality of the
material.

All cyclotron resonance measurements were performed at 4.2 K. A
Fourier transform spectrometer with light pipe optics was used in
combination with a composite Si bolometer for the detection of the
far-infrared (FIR) magneto-transmission. The backsides of samples
3,5,8, and 10 were wedged to $\sim$9$^\circ$ to reduce
Fabry-Perrot interferences . The $B$-field was applied normal to
the 2D layer and the density of the 2DES was determined \textit{in
situ} via the Shubnikov-de Haas effect. The electron densities in
samples 1-6 could be persistently increased (up to $30 \%$)
through illuminations with a blue LED for 20 - 60 seconds.
\begin{figure}
\includegraphics{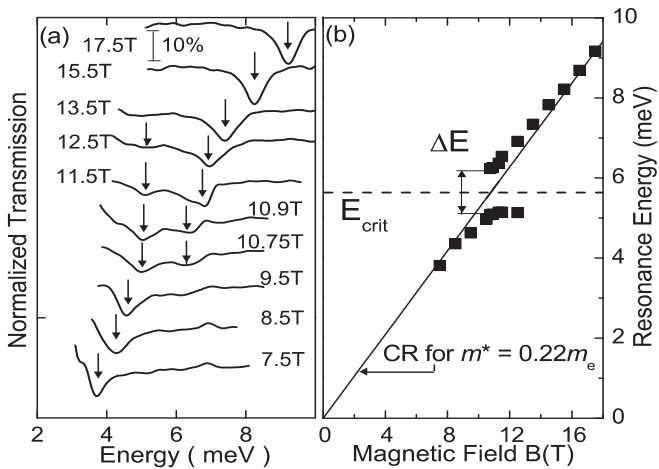}
\vspace{-0.2in} \caption{\label{Fig.1} (a) Far infrared
transmission data on a 2DES of
 density $1.14 \times 10^{12}$ cm$^{-2}$ in AlGaN/GaN (sample 3) for various magnetic
 fields, $B$, normal to the 2DES. All data are normalized to the transmission
 at $B$=0. Traces are offset vertically for clarity. A
 $10\%$ transmission loss is indicated as a vertical bar.
 Sharp CR lines are observed for $B<9$T and $B>15$T. In the intermediate
 field regime large splittings  occur. (b) Peak positions of transmission minima of
(a) as a function of magnetic field. High and low field data
follow the CR with an effective mass of $m^*$=0.22$m_e$. Around
11T an apparent level anti-crossing occurs with a critical energy
E$_\mathrm{crit}$ =5.6meV.}
\end{figure}

Fig.1a shows a representative set of transmission spectra for
sample 3 in different magnetic fields. All spectra are taken with
0.24 meV resolution and are normalized to the spectrum obtained at
0T. For clarity the data are vertically offset. For fields $B>15$T
and $B<8$T  singular sharp resonance dips are observed in
transmission. They represent the characteristic CR of the
electrons. The 2D nature of the carrier motion is verified  by
observing the expected $1/\cos\theta$ shift of the resonances
under tilted $B$ field. These high and low field resonance
positions are linearly dependent on $B$, are consistent with each
other, and yield an electron effective mass of $m^*=0.22m_e$ which
is very close to the literature value \cite{Knap96}.

While the CR is well behaved in the high-field and low-field
regimes, at intermediate fields (8T$<B<$15T) it deviates
considerably from a simple linear relationship. As the field is
reduced below $B\sim$17T the line-width increases gradually and
develops a separate transmission dip near 13T. For example, at
$B=12.5$T a strong resonance exists near 6.9meV and a weaker one
at 5.1meV.  Lowering the field further, the high-energy resonance
gradually broadens and weakens, while the low-energy resonance
gains in strength. At $B\sim$11.3T, both resonances are of equal
strength. The integrated intensity of the combined resonances
remains comparable to those of the single CR line for $B>15$T. As
$B$ moves towards 8T, the low energy resonance continues to grow
and develops again into a single CR. For lower fields, the
resonance dips regain their sharpness and move linearly with $B$
to smaller energies.

Fig.1b shows the position of the resonance energies of sample 3
versus magnetic field. The solid line is a fit to the high and low
field data. Between 10T and 14T we observe a marked deviation from
a linear dependence together with the appearance of a second
resonance for each magnetic field. The two resonance branches are
separated by a gap, $\Delta$E, of $\sim$1.2meV. The midpoint of
this gap, E$_\mathrm{crit}$, is $\sim$5.6 meV. The splitting
resembles a level anti-crossing between the CR of the system and
some other resonance at $\sim$5.6 meV. However, this ``other
resonance'' is not observed outside of the crossing regime. In
particular, it is not optically active at $B=0$.

A clear splitting in the CR is observed in seven different samples
and a distinct line broadening is observed in three others. Figure
2 shows the dependence of E$_\mathrm{crit}$ and $\Delta$E on the
2D density, $n_{2D}$, for all 10 samples. For specimens in which
the splitting is not resolved, we take E$_\mathrm{crit}$  as the
energy of the maximally broadened line. The density dependence of
the CR splitting, considering the scatter of the data, is
approximately constant, $\Delta$E$\sim$1.5meV. Beyond our own
data, Wang \textit{et al.} \cite{Wang96} reported a broadening of
the transmission dip near 8.6meV. We obtained this specimen,
determined its density  and included these data in our graph.
Their position indicates that the same phenomenon is at work.
Other CR experiments on AlGaN$/$GaN heterostructures do not
observe a splitting nor a broadening \cite{Knap96,Knap97,ZFLi02}.
These experiments were performed on low mobility and very
high-density specimens ($\sim$ 3.1$\times$10$^{12}$ $<$
$n_{2D}$$<$$\sim$ 18$\times$ 10$^{12}$ cm$^{-2}$). Judging from
Fig. 2, the level anti-crossing position in most of these
experiments exceeded the maximum field employed. In others the
splitting may have been missed because of much wider CR
line-widths and sparse data sets.

The splitting of the CR line in our samples appears to be the
result of the coupling of the 2DES CR with another resonance of
the system with an energy in the 5~-~12~meV range. Interaction
between the 2D electrons and bulk or interface phonons cannot be
the cause of the CR splitting since all optical phonon energies in
AlGaN/GaN are above 65meV \cite{Azuhata95,Gleize99}. The common
splitting due to a coincidence of the CR with the intersubband
separation of the 2DES can also be ruled out: (a) the calculated
subband separation is $\sim$23meV for $n_{2D}\sim1\times10^{12}$
cm$^{-2}$ \cite{LHsu97} which is much bigger than the observed
$\sim$6meV; (b) such interaction occurs only at finite angles,
whereas we observe the splitting for $\theta$=0$^\circ$ and find
no observable angular dependence. We can also safely rule out a
coincidence of opposite spin states in neighboring Landau levels
\cite{Falko92}. For our material system such a coincidence would
occur for $\theta \sim$77$^\circ$. The observed splitting cannot
be due to a coupling of the 2DES with any 3D plasmon in any of the
layers since there are very few carriers in the bulk. For example,
for $n_{2D}\sim$1$\times 10^{12}$ cm$^{-2}$, the splitting occurs
near 6meV which requires a carrier density of $\sim$10$^{17}$
cm$^{-3}$ in a 3D plasmon, whereas C-V measurements of the bulk
set those densities  to $<\sim$ 5$\times$10$^{14}$ cm$^{-3}$
\cite{Manfra00}.
\begin{table}
  \centering
  \caption{Parameters of all AlGaN/GaN 2DES samples used in our
  experiments. $n_{2D}$ in units of 10$^{11}$cm$^{-2}$ and the mobility, $\mu$, in $10^3$ cm$^2$/Vs. E$_\mathrm{crit}$ and
  $\Delta$E are in meV.
  The values in parentheses are estimated from deconvolutions.}\label{table}
\newcolumntype{Y}{>{\centering\arraybackslash}X}%
\begin{tabularx}{\linewidth}{|Y|Y|Y|Y|Y||Y|Y|Y|Y|Y|}
\hline $\#$ & $n_{2D}$ & $\mu$ & E$_\mathrm{crit}$ & $\Delta$E & $\#$ & $n_{2D}$ & $\mu$ & E$_\mathrm{crit}$ & $\Delta$E\\
\hline 1 & 9.8 & 16 & 4.9 & 1.6 & 6 & 12.6 & 17 & 6.5 & 1.3 \\
 \hline 2 & 11.2 & 15 & 5.6 & 1.7 & 7 & 19.0 & 16 & 7.3 & (0.8) \\
\hline 3 & 11.4 & 16 & 5.6 & 1.2 & 8 & 23.0 & 18 & 8.9 & 1.7 \\
\hline 4 & 11.9 & 20 & 5.8 & 1.7 & 9 & 35.8 & 8 & 10.4 & (1.6) \\
\hline 5 & 12.3 & 18 & 6.1 & 1.4 & 10 & 36.1 & 19 & 12.2 & (2.0) \\
\hline
\end{tabularx}
\end{table}

The closest resemblance to our observations is found in a previous
report by Schlesinger \textit{et al.} \cite{Schlesinger} who
observed low-energy splittings in the CR of AlGaAs/GaAs
heterostructures covering a density range of $1$- $4\times
10^{11}$ cm$^{-2}$. Their data remain unexplained
(e.g.Ref.\cite{Zhao95}). Several observations of
Ref.\cite{Schlesinger} are similar to ours: (a) the CR splitting
is not affected by small tilt angles, (b) no absorption is seen in
the $B=0$T spectrum at the critical energy, (c) at the critical
$B$-field two resonances of roughly equal width and strength
appear, and (d) the energy, E$_\mathrm{crit}$, at which a
broadening/splitting is observed is proportional to
$\sqrt{n_{2D}}$  (see Fig.2). While the similarities in these
observations suggest a common origin of the CR splittings in
AlGaAs/GaAs and our AlGaN/GaN interface, there are some
differences. In AlGaAs/GaAs heterostructures the splitting appears
``abruptly" whereas our AlGaN/GaN data show a gradual evolution of
the splitting. Moreover, the separation between the split lines at
the critical $B$-field is $\sim5\%$ of the CR energy in the
AlGaAs/GaAs case, much smaller than the $\sim 20\%$ seen in our
AlGaN/GaN heterostructures.

The most important difference is a pronounced deviation of our
data from a universality proposed by Ref.\cite{Schlesinger}. By
including previous results of Wilson \textit{et al.}
\cite{Wilson80} and Kennedy \textit{et al.} \cite{Kennedy77} on CR
line broadening in Si, Schlesinger \textit{et al.} infer that the
critical energy follows a universal relationship
E$_\mathrm{crit}$=$\sqrt{n_{2D}}e^2/\epsilon$, where $\epsilon$ is
the dielectric constant. While our data also seem to follow a
$\sqrt{n_{2D}}$ dependence they deviate by a factor of $\sim$2.5
from the AlGaAs/GaAs and Si case (see Fig. 2). Obvious differences
in $\epsilon$ and $m^*$ between these materials cannot resolve the
discrepancy. In fact, $\epsilon$  and $m^*$ of Si and GaN are
within $15\%$ of each other and yet there exists a factor of 2.5
discrepancy in Fig. 2. A plot of E$_\mathrm{crit}$ vs. filling
factor, $\nu$, generates no apparent relationship. On the other
hand, in a plot of $B_\mathrm{crit}$ vs. $n_{2D}$ the AlGaAs/GaAs
and AlGaN/GaN data fall onto the same line. However, now the Si
data deviate by a factor of $\sim$3.
\begin{figure}
\includegraphics{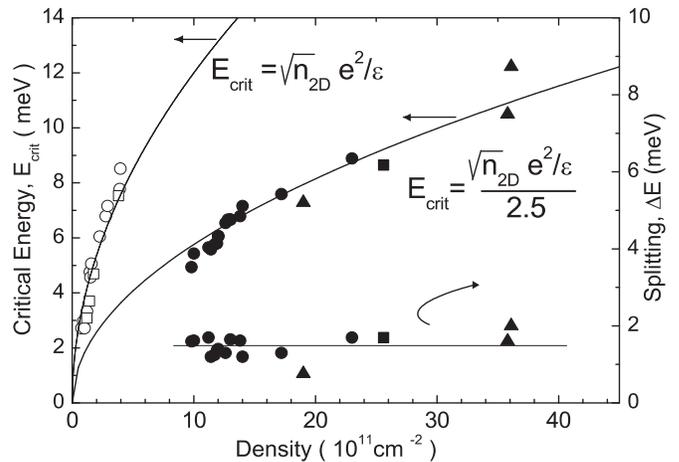}
\caption{\label{Fig.2} Critical energy, E$_\mathrm{crit}$, (left
scale) and splitting, $\Delta$E, (right scale) versus electron
density. All AlGaN/GaN samples of Table I are shown plus data from
runs in which the density was increased (up to $30\%$) by light.
Data from clearly resolved CR splittings are indicated by filled
circles, broadenings are shown as filled triangles. The full
squares are data from Ref. \cite{Wang96}. Open circles and open
squares refer to AlGaAs/GaAs data (Ref.\cite{Schlesinger}) and Si
data (Ref.\cite{Wilson80,Kennedy77}) respectively. All data follow
a $\sqrt{n_{2D}}$ behavior, however with very different
prefactors. The data set at the bottom ($\Delta$E) shows the
magnitude of the CR splitting at $B_\mathrm{crit}$ as a function
of density for all AlGaN/GaN samples. In the case of line
broadening (triangles and square) the ``splitting" was estimated
from a deconvolution of the broadened CR line.}
\end{figure}

As to the origin of the CR splitting, Schlesinger \textit{et al.}
propose an \textit{ad hoc} model. The splitting is conjectured to
result from a softening of the large $q \sim 2/l_0$ magnetoroton
mode leading to a degeneracy with the CR at $q=0$. The plasmon
wave vector is denoted as $q$ and  $l_0$=$\sqrt{\hbar c/eB}$ is
the magnetic length. Disorder breaks translation invariance,
couples both modes, and causes the splitting. For this reason, the
splitting is absent in very high-mobility specimens such as
AlGaAs/GaAs with $\mu>10^6$ cm$^2/$Vsec. The origin of the
softening of the magnetoroton and its $B$-dependence remain
speculative in Ref.\cite{Schlesinger}. Kallin and Halperin
\cite{Kallin85} are among the first to address the
field-dependence of the magneto-plasmon dispersion. Their analysis
is limited to high magnetic fields where the cyclotron energy
$\hbar\omega_c>>E_c$ , the Coulomb energy. Although in the
experiments $E_c\approx$ $\hbar$$\omega_c$, the authors provide
some general remarks about the CR splitting in AlGaAs/GaAs.
According to their calculation the magnetoroton minimum never
approaches the CR energy but always exceeds it. Even if the
magnetoroton were to cross the CR due to some higher order
interaction it would do so in the wrong direction. The roton
minimum is expected to move downward in energy with decreasing
field, in contrast to experiment, which requires the minimum to
move upward in energy with decreasing field. A calculation by
MacDonald \cite{MacDonald85}, which includes higher order effects
on the magnetoplasmon dispersion, also concludes that the
corrections are too small to even bring the minimum into resonance
with the CR. A recent calculation on interactions between
magnetoplasmons by Cheng \cite{Cheng94} asserts that the
magnetoroton minimum actually can cross the CR energy. Cheng
calculates dispersion relations only in the absence of disorder
and does not derive actual splittings of the CR line. Furthermore,
as previously argued by Kallin and Halperin, the direction of
crossing of the magnetoroton minimum, calculated by Cheng
including these higher order effects, is again inconsistent with
experiment (upward vs. downward in energy as a function of $B$).
All previous references restrict their calculations to integral
values of filling factor, $\nu$. Oji and MacDonald \cite{Oji86}
consider the case of arbitrary filling factors but find no
particularly strong dependence of the roton minimum energy on
$\nu$. Instead, they propose that a combined action of spin
density mode (largely below CR) and charge density mode (largely
above CR) may be responsible for the CR line splitting. However,
the coupling of spin and charge to the CR mode are very different
in strength and runs counter to the observation of a simple level
anti-crossing behavior. There are other attempts at interpreting
the AlGaAs/GaAs CR data using a memory function approach. Gold's
calculations \cite{Gold90}, performed in the small-$q$ limit can
only account for a broadening but not for a splitting of the CR
line. The calculations by Hu and O'Connell \cite{Hu88}, also based
on memory functions, remarkably, generate such a splitting.
However, the center of gravity of the combined lines (see Fig. 3)
always tend to reside above $\hbar\omega_c$ in conflict with
experiment. Furthermore, extrapolating these results, equal
amplitude of the peaks would occur at $B \sim 8$T
(E$_\mathrm{crit} \sim 5.3$meV) which deviates considerably from
the E$_\mathrm{crit} \sim 12$meV of Ref.\cite{Schlesinger}.

At this stage, we do not understand the splitting in the CR of 2D
electrons in AlGaN/GaN heterostructures. The splitting is
enormous, reaching up to $\sim20\%$ of the CR energy. Its
unambiguous observation in a second material system establishes
the splitting as a general 2D phenomenon rather than being
peculiar to just AlGaAs/GaAs.

Theoretical models that address the splitting in AlGaAs/GaAs
appeal to a softening of the large wave vector magnetoplasmon
mode, which mixes into the CR due to disorder. Yet the origin,
$B$-field dependence, filling factor dependence, magnitude, and
even the exact mode responsible for the mixing remain unresolved.
At the present level of understanding of 2D electron dynamics it
is remarkable that such a huge effect on one of the most
fundamental excitations of a solid, the cyclotron resonance of
electrons, remains obscure. Our data on the new AlGaN/GaN
heterostructures clearly highlights the need for a detailed
evaluation of the magnetoplasmon mode. Our observed density
dependence, which differs significantly from an earlier
conjecture, should provide a means to differentiate between
various theoretical models.

\begin{acknowledgments}
We are grateful for helpful discussions with A. Millis, A.
MacDonald, A. Pinczuk, C. F. Hirjibehedin, L. N. Pfeiffer and K.
W. West. A portion of the work was performed at the National High
Magnetic Field Laboratory, which is supported by NSF Cooperative
Agreement No. DMR-9527035 and by the State of Florida. Financial
support from the W. M. Keck Foundation is gratefully acknowledged.
\end{acknowledgments}


\begin{thebibliography}{24}
\expandafter\ifx\csname
natexlab\endcsname\relax\def\natexlab#1{#1}\fi
\expandafter\ifx\csname bibnamefont\endcsname\relax
  \def\bibnamefont#1{#1}\fi
\expandafter\ifx\csname bibfnamefont\endcsname\relax
  \def\bibfnamefont#1{#1}\fi
\expandafter\ifx\csname citenamefont\endcsname\relax
  \def\citenamefont#1{#1}\fi
\expandafter\ifx\csname url\endcsname\relax
  \def\url#1{\texttt{#1}}\fi
\expandafter\ifx\csname
urlprefix\endcsname\relax\def\urlprefix{URL }\fi
\providecommand{\bibinfo}[2]{#2}
\providecommand{\eprint}[2][]{\url{#2}}

\bibitem[{\citenamefont{Dresselhaus et~al.}(1953)\citenamefont{Dresselhaus,
  Kip, and Kittel}}]{Kittel53}
\bibinfo{author}{\bibfnamefont{G.}~\bibnamefont{Dresselhaus}},
  \bibinfo{author}{\bibfnamefont{A.~F.} \bibnamefont{Kip}}, \bibnamefont{and}
  \bibinfo{author}{\bibfnamefont{C.}~\bibnamefont{Kittel}},
  \bibinfo{journal}{Phys. Rev.} \textbf{\bibinfo{volume}{92}},
  \bibinfo{pages}{827} (\bibinfo{year}{1953}).

\bibitem[{\citenamefont{McCombe and Petrou}(1994)}]{McCombe1994}
\bibinfo{author}{\bibfnamefont{B.~D.} \bibnamefont{McCombe}} \bibnamefont{and}
  \bibinfo{author}{\bibfnamefont{A.}~\bibnamefont{Petrou}},
  \emph{\bibinfo{title}{Handbook on Semiconductors,}} ed. T.S. Moss and M.
  Balkanski (\bibinfo{publisher}{Elsevier Science B.V.},
  \bibinfo{year}{1994}), vol.~\bibinfo{volume}{2}, chap.~\bibinfo{chapter}{6},
  p. \bibinfo{pages}{297}.

\bibitem[{\citenamefont{Wang et~al.}(1996)\citenamefont{Wang, Kaplan, Ng,
  Doverspike, Gaskill, Ikedo, Akasaki, and Amono}}]{Wang96}
\bibinfo{author}{\bibfnamefont{Y.~J.} \bibnamefont{Wang}},
  \bibinfo{author}{\bibfnamefont{R.}~\bibnamefont{Kaplan}},
  \bibinfo{author}{\bibfnamefont{H.}~\bibnamefont{Ng}},
  \bibinfo{author}{\bibfnamefont{K.}~\bibnamefont{Doverspike}},
  \bibinfo{author}{\bibfnamefont{D.}~\bibnamefont{Gaskill}},
  \bibinfo{author}{\bibfnamefont{T.}~\bibnamefont{Ikedo}},
  \bibinfo{author}{\bibfnamefont{I.}~\bibnamefont{Akasaki}}, \bibnamefont{and}
  \bibinfo{author}{\bibfnamefont{H.}~\bibnamefont{Amono}}, \bibinfo{journal}{J.
  Appl. Phys.} \textbf{\bibinfo{volume}{79}}, \bibinfo{pages}{8007}
  (\bibinfo{year}{1996}).

\bibitem[{\citenamefont{Knap et~al.}(1996)\citenamefont{Knap, Alause, Bluet,
  Camassel, Young, Khan, Chen, Huant, and Shur}}]{Knap96}
\bibinfo{author}{\bibfnamefont{W.}~\bibnamefont{Knap}},
  \bibinfo{author}{\bibfnamefont{H.}~\bibnamefont{Alause}},
  \bibinfo{author}{\bibfnamefont{J.}~\bibnamefont{Bluet}},
  \bibinfo{author}{\bibfnamefont{J.}~\bibnamefont{Camassel}},
  \bibinfo{author}{\bibfnamefont{J.}~\bibnamefont{Young}},
  \bibinfo{author}{\bibfnamefont{M.~A.} \bibnamefont{Khan}},
  \bibinfo{author}{\bibfnamefont{Q.}~\bibnamefont{Chen}},
  \bibinfo{author}{\bibfnamefont{S.}~\bibnamefont{Huant}}, \bibnamefont{and}
  \bibinfo{author}{\bibfnamefont{M.}~\bibnamefont{Shur}},
  \bibinfo{journal}{Solid State Comm.} \textbf{\bibinfo{volume}{99}},
  \bibinfo{pages}{195} (\bibinfo{year}{1996}).

\bibitem[{\citenamefont{Knap et~al.}(1997)\citenamefont{Knap, Contreas, Alause,
  Skierbiszewski, Camassel, Dyakonov, Yang, Khan, Chen, Yang et~al.}}]{Knap97}
\bibinfo{author}{\bibfnamefont{W.}~\bibnamefont{Knap}},
  \bibinfo{author}{\bibfnamefont{S.}~\bibnamefont{Contreas}},
  \bibinfo{author}{\bibfnamefont{H.}~\bibnamefont{Alause}},
  \bibinfo{author}{\bibfnamefont{C.}~\bibnamefont{Skierbiszewski}},
  \bibinfo{author}{\bibfnamefont{J.}~\bibnamefont{Camassel}},
  \bibinfo{author}{\bibfnamefont{M.}~\bibnamefont{Dyakonov}},
  \bibinfo{author}{\bibfnamefont{J.}~\bibnamefont{Yang}},
  \bibinfo{author}{\bibfnamefont{M.~A.} \bibnamefont{Khan}},
  \bibinfo{author}{\bibfnamefont{Q.}~\bibnamefont{Chen}},
  \bibinfo{author}{\bibfnamefont{J.}~\bibnamefont{Yang}}, \bibnamefont{et~al.},
  \bibinfo{journal}{Appl. Phys. Lett.} \textbf{\bibinfo{volume}{70}},
  \bibinfo{pages}{2123} (\bibinfo{year}{1997}).

\bibitem[{\citenamefont{Li et~al.}(2002)\citenamefont{Li, Lu, Shen, Holland,
  Hu, Heitmann, Shen, and Zheng}}]{ZFLi02}
\bibinfo{author}{\bibfnamefont{Z.-F.} \bibnamefont{Li}},
  \bibinfo{author}{\bibfnamefont{W.}~\bibnamefont{Lu}},
  \bibinfo{author}{\bibfnamefont{S.}~\bibnamefont{Shen}},
  \bibinfo{author}{\bibfnamefont{S.}~\bibnamefont{Holland}},
  \bibinfo{author}{\bibfnamefont{C.}~\bibnamefont{Hu}},
  \bibinfo{author}{\bibfnamefont{D.}~\bibnamefont{Heitmann}},
  \bibinfo{author}{\bibfnamefont{B.}~\bibnamefont{Shen}}, \bibnamefont{and}
  \bibinfo{author}{\bibfnamefont{Y.}~\bibnamefont{Zheng}},
  \bibinfo{journal}{Appl. Phys. Lett.} \textbf{\bibinfo{volume}{80}},
  \bibinfo{pages}{431} (\bibinfo{year}{2002}).

\bibitem[{\citenamefont{Ambacher et~al.}(1999)\citenamefont{Ambacher, Smart,
  Shealy, Weimann, Chu, M.Murphy, Schaff, Eastman, Dimitrov, Wittmer
  et~al.}}]{Ambacher99}
\bibinfo{author}{\bibfnamefont{O.}~\bibnamefont{Ambacher}},
  \bibinfo{author}{\bibfnamefont{J.}~\bibnamefont{Smart}},
  \bibinfo{author}{\bibfnamefont{J.}~\bibnamefont{Shealy}},
  \bibinfo{author}{\bibfnamefont{N.}~\bibnamefont{Weimann}},
  \bibinfo{author}{\bibfnamefont{K.}~\bibnamefont{Chu}},
  \bibinfo{author}{\bibnamefont{M.Murphy}},
  \bibinfo{author}{\bibfnamefont{W.}~\bibnamefont{Schaff}},
  \bibinfo{author}{\bibfnamefont{L.}~\bibnamefont{Eastman}},
  \bibinfo{author}{\bibfnamefont{R.}~\bibnamefont{Dimitrov}},
  \bibinfo{author}{\bibfnamefont{L.}~\bibnamefont{Wittmer}},
  \bibnamefont{et~al.}, \bibinfo{journal}{J. Appl. Phys.}
  \textbf{\bibinfo{volume}{85}}, \bibinfo{pages}{3222} (\bibinfo{year}{1999}).

\bibitem[{\citenamefont{Smorchkova et~al.}(1999)\citenamefont{Smorchkova,
  Elsass, Ibbetson, Vetury, Heying, Fini, Haus, DenBaars, Speck, and
  Mishra}}]{Smorchkova99}
\bibinfo{author}{\bibfnamefont{I.}~\bibnamefont{Smorchkova}},
  \bibinfo{author}{\bibfnamefont{C.}~\bibnamefont{Elsass}},
  \bibinfo{author}{\bibfnamefont{J.}~\bibnamefont{Ibbetson}},
  \bibinfo{author}{\bibfnamefont{R.}~\bibnamefont{Vetury}},
  \bibinfo{author}{\bibfnamefont{B.}~\bibnamefont{Heying}},
  \bibinfo{author}{\bibfnamefont{P.}~\bibnamefont{Fini}},
  \bibinfo{author}{\bibfnamefont{E.}~\bibnamefont{Haus}},
  \bibinfo{author}{\bibfnamefont{S.}~\bibnamefont{DenBaars}},
  \bibinfo{author}{\bibfnamefont{J.}~\bibnamefont{Speck}}, \bibnamefont{and}
  \bibinfo{author}{\bibfnamefont{U.}~\bibnamefont{Mishra}},
  \bibinfo{journal}{J. Appl. Phys.} \textbf{\bibinfo{volume}{86}},
  \bibinfo{pages}{4520} (\bibinfo{year}{1999}).

\bibitem[{\citenamefont{Manfra et~al.}(2000)\citenamefont{Manfra, Pfeiffer,
  West, Stormer, Baldwin, Hsu, Lang, and Molnar}}]{Manfra00}
\bibinfo{author}{\bibfnamefont{M.~J.} \bibnamefont{Manfra}},
  \bibinfo{author}{\bibfnamefont{L.~N.} \bibnamefont{Pfeiffer}},
  \bibinfo{author}{\bibfnamefont{K.~W.} \bibnamefont{West}},
  \bibinfo{author}{\bibfnamefont{H.~L.} \bibnamefont{Stormer}},
  \bibinfo{author}{\bibfnamefont{K.~W.} \bibnamefont{Baldwin}},
  \bibinfo{author}{\bibfnamefont{J.~W.~P.} \bibnamefont{Hsu}},
  \bibinfo{author}{\bibfnamefont{D.~V.} \bibnamefont{Lang}}, \bibnamefont{and}
  \bibinfo{author}{\bibfnamefont{R.~J.} \bibnamefont{Molnar}},
  \bibinfo{journal}{Appl. Phys. Lett.} \textbf{\bibinfo{volume}{77}},
  \bibinfo{pages}{2888} (\bibinfo{year}{2000}).

\bibitem[{\citenamefont{Manfra et~al.}(2002)\citenamefont{Manfra, Weimann, Hsu,
  Pfeiffer, West, Syed, Stormer, Pan, Lang, Chu et~al.}}]{Manfra02}
\bibinfo{author}{\bibfnamefont{M.~J.} \bibnamefont{Manfra}},
  \bibinfo{author}{\bibfnamefont{N.~G.} \bibnamefont{Weimann}},
  \bibinfo{author}{\bibfnamefont{J.~W.~P.} \bibnamefont{Hsu}},
  \bibinfo{author}{\bibfnamefont{L.~N.} \bibnamefont{Pfeiffer}},
  \bibinfo{author}{\bibfnamefont{K.~W.} \bibnamefont{West}},
  \bibinfo{author}{\bibfnamefont{S.}~\bibnamefont{Syed}},
  \bibinfo{author}{\bibfnamefont{H.~L.} \bibnamefont{Stormer}},
  \bibinfo{author}{\bibfnamefont{W.}~\bibnamefont{Pan}},
  \bibinfo{author}{\bibfnamefont{D.~V.} \bibnamefont{Lang}},
  \bibinfo{author}{\bibfnamefont{S.~N.} \bibnamefont{Chu}},
  \bibnamefont{et~al.}, \bibinfo{journal}{J. Appl. Phys}
  \textbf{\bibinfo{volume}{92}}, \bibinfo{pages}{338} (\bibinfo{year}{2002}).

\bibitem[{\citenamefont{Azuhata et~al.}(1995)\citenamefont{Azuhata, Sota,
  Suzuki, and Nakamura}}]{Azuhata95}
\bibinfo{author}{\bibfnamefont{T.}~\bibnamefont{Azuhata}},
  \bibinfo{author}{\bibfnamefont{T.}~\bibnamefont{Sota}},
  \bibinfo{author}{\bibfnamefont{K.}~\bibnamefont{Suzuki}}, \bibnamefont{and}
  \bibinfo{author}{\bibfnamefont{S.}~\bibnamefont{Nakamura}},
  \bibinfo{journal}{J. Phys.: Condens. Matter} \textbf{\bibinfo{volume}{7}},
  \bibinfo{pages}{L129} (\bibinfo{year}{1995}).

\bibitem[{\citenamefont{Gleize et~al.}(1999)\citenamefont{Gleize, Renucci,
  Frandon, and Demangeot}}]{Gleize99}
\bibinfo{author}{\bibfnamefont{J.}~\bibnamefont{Gleize}},
  \bibinfo{author}{\bibfnamefont{M.~A.} \bibnamefont{Renucci}},
  \bibinfo{author}{\bibfnamefont{J.}~\bibnamefont{Frandon}}, \bibnamefont{and}
  \bibinfo{author}{\bibfnamefont{F.}~\bibnamefont{Demangeot}},
  \bibinfo{journal}{Phys. Rev. B} \textbf{\bibinfo{volume}{60}},
  \bibinfo{pages}{15985} (\bibinfo{year}{1999}).

\bibitem[{\citenamefont{L.Hsu and W.Walukiewicz}(1997)}]{LHsu97}
\bibinfo{author}{\bibnamefont{L.Hsu}} \bibnamefont{and}
  \bibinfo{author}{\bibnamefont{W.Walukiewicz}}, \bibinfo{journal}{Phys. Rev.
  B} \textbf{\bibinfo{volume}{56}}, \bibinfo{pages}{1520}
  (\bibinfo{year}{1997}).

\bibitem[{\citenamefont{Falko}(1992)}]{Falko92}
\bibinfo{author}{\bibfnamefont{V.~I.} \bibnamefont{Falko}},
  \bibinfo{journal}{Phys. Rev. B} \textbf{\bibinfo{volume}{46}},
  \bibinfo{pages}{4320} (\bibinfo{year}{1992}).

\bibitem[{\citenamefont{Schlesinger et~al.}(1984)\citenamefont{Schlesinger,
  Allen, Jr., Hwang, Platzman, and Tzoar}}]{Schlesinger}
\bibinfo{author}{\bibfnamefont{Z.}~\bibnamefont{Schlesinger}},
  \bibinfo{author}{\bibfnamefont{S.~J.} \bibnamefont{Allen}},
  \bibinfo{author}{\bibnamefont{Jr.}}, \bibinfo{author}{\bibfnamefont{J.~C.~M.}
  \bibnamefont{Hwang}}, \bibinfo{author}{\bibfnamefont{P.~M.}
  \bibnamefont{Platzman}}, \bibnamefont{and}
  \bibinfo{author}{\bibfnamefont{N.}~\bibnamefont{Tzoar}},
  \bibinfo{journal}{Phys. Rev. B} \textbf{\bibinfo{volume}{30}},
  \bibinfo{pages}{435} (\bibinfo{year}{1984}).

\bibitem[{\citenamefont{Zhao et~al.}(1995)\citenamefont{Zhao, Tsui, Santos,
  Shayegan, Ghanbari, Antoniadis, and Smith}}]{Zhao95}
\bibinfo{author}{\bibfnamefont{Y.}~\bibnamefont{Zhao}},
  \bibinfo{author}{\bibfnamefont{D.~C.} \bibnamefont{Tsui}},
  \bibinfo{author}{\bibfnamefont{M.~B.} \bibnamefont{Santos}},
  \bibinfo{author}{\bibfnamefont{M.}~\bibnamefont{Shayegan}},
  \bibinfo{author}{\bibfnamefont{R.~A.} \bibnamefont{Ghanbari}},
  \bibinfo{author}{\bibfnamefont{D.~A.} \bibnamefont{Antoniadis}},
  \bibnamefont{and} \bibinfo{author}{\bibfnamefont{H.~I.} \bibnamefont{Smith}},
  \bibinfo{journal}{Phys. Rev. B} \textbf{\bibinfo{volume}{51}},
  \bibinfo{pages}{13174} (\bibinfo{year}{1995}).

\bibitem[{\citenamefont{Wilson et~al.}(1980)\citenamefont{Wilson, Allen, Jr.,
  and Tsui}}]{Wilson80}
\bibinfo{author}{\bibfnamefont{B.~A.} \bibnamefont{Wilson}},
  \bibinfo{author}{\bibfnamefont{S.~J.} \bibnamefont{Allen}},
  \bibinfo{author}{\bibnamefont{Jr.}}, \bibnamefont{and}
  \bibinfo{author}{\bibfnamefont{D.~C.} \bibnamefont{Tsui}},
  \bibinfo{journal}{Phys. Rev. Lett.} \textbf{\bibinfo{volume}{44}},
  \bibinfo{pages}{479} (\bibinfo{year}{1980}); \bibinfo{journal}{Phys. Rev. B} \textbf{\bibinfo{volume}{24}},
  \bibinfo{pages}{5887} (\bibinfo{year}{1981}).

\bibitem[{\citenamefont{Kennedy et~al.}(1977)\citenamefont{Kennedy, Wagner,
  McCombe, and Tsui}}]{Kennedy77}
\bibinfo{author}{\bibfnamefont{T.~A.} \bibnamefont{Kennedy}},
  \bibinfo{author}{\bibfnamefont{R.~J.} \bibnamefont{Wagner}},
  \bibinfo{author}{\bibfnamefont{B.~D.} \bibnamefont{McCombe}},
  \bibnamefont{and} \bibinfo{author}{\bibfnamefont{D.~C.} \bibnamefont{Tsui}},
  \bibinfo{journal}{Solid State Commun.} \textbf{\bibinfo{volume}{21}},
  \bibinfo{pages}{459} (\bibinfo{year}{1977}).

\bibitem[{\citenamefont{Kallin and Halperin}(1985)}]{Kallin85}
\bibinfo{author}{\bibfnamefont{C.}~\bibnamefont{Kallin}} \bibnamefont{and}
  \bibinfo{author}{\bibfnamefont{B.~I.} \bibnamefont{Halperin}},
  \bibinfo{journal}{Phys. Rev. B} \textbf{\bibinfo{volume}{31}},
  \bibinfo{pages}{3635} (\bibinfo{year}{1985}).

\bibitem[{\citenamefont{MacDonald}(1985)}]{MacDonald85}
\bibinfo{author}{\bibfnamefont{A.~H.} \bibnamefont{MacDonald}},
  \bibinfo{journal}{J. Phys. C} \textbf{\bibinfo{volume}{18}},
  \bibinfo{pages}{1003} (\bibinfo{year}{1985}).

\bibitem[{\citenamefont{Cheng}(1994)}]{Cheng94}
\bibinfo{author}{\bibfnamefont{S.-C.} \bibnamefont{Cheng}},
  \bibinfo{journal}{Phys. Rev. B} \textbf{\bibinfo{volume}{49}},
  \bibinfo{pages}{4703} (\bibinfo{year}{1994}).

\bibitem[{\citenamefont{Oji and MacDonald}(1986)}]{Oji86}
\bibinfo{author}{\bibfnamefont{H.~C.~A.} \bibnamefont{Oji}} \bibnamefont{and}
  \bibinfo{author}{\bibfnamefont{A.~H.} \bibnamefont{MacDonald}},
  \bibinfo{journal}{Phys. Rev. B} \textbf{\bibinfo{volume}{33}},
  \bibinfo{pages}{3810} (\bibinfo{year}{1986}).

\bibitem[{\citenamefont{Gold}(1990)}]{Gold90}
\bibinfo{author}{\bibfnamefont{A.}~\bibnamefont{Gold}}, \bibinfo{journal}{Phys.
  Rev. B} \textbf{\bibinfo{volume}{41}}, \bibinfo{pages}{3608}
  (\bibinfo{year}{1990}).

\bibitem[{\citenamefont{Hu and O'Connell}(1988)}]{Hu88}
\bibinfo{author}{\bibfnamefont{G.~Y.} \bibnamefont{Hu}} \bibnamefont{and}
  \bibinfo{author}{\bibfnamefont{R.~F.} \bibnamefont{O'Connell}},
  \bibinfo{journal}{Phys. Rev. B} \textbf{\bibinfo{volume}{37}},
  \bibinfo{pages}{10391} (\bibinfo{year}{1988}).

\end{thebibliography}

\end{document}